\documentclass[pra,aps,amsmath,twocolumn,showpacs,floatfix]{revtex4-1}
\usepackage{amssymb}
\usepackage{graphicx,xcolor}
\usepackage{epigraph}
\usepackage{csquotes}
\usepackage{amsmath,amssymb,stmaryrd}
\usepackage{amsmath,graphicx}
\begin{document}
\def\la{{\langle}}
\def\u{\hat U}
\def\U{\hat U}
\def\B{\hat B}
\def\C{\hat C}
\def\Q{\hat Q}
\def\D{\hat D}
\def\e{\enquote}
\def\fb{\overline F}
\def\wb{\overline W}
\def\nl{\newline}
\def\h{\hat H}
\def\lm{\lambda}
\def\lmu{\underline\lambda}
\def\q{\quad}
\def\t{\tau}
\def\l{\ell}
\def\n{\\ \nonumber}
\def\ra{{\rangle}}
\def\Ep{{\mathcal{E}}}
\def\omga{{\epsilon}}
\def\t{{\tau}}
\def\h{\hat{H}}
\title{Wigner's friend, Feynman's paths and material records}
%
%
\author {A. Matzkin$^{a}$}
\author {D. Sokolovski$^{b,c}$}
\affiliation{$^a$ Laboratoire de Physique Th\'eorique et Mod\'elisation, CNRS Unit\'e 8089, CY Cergy Paris Universit\'e,
95302 Cergy-Pontoise cedex, France}
\affiliation{$^b$ Departmento de Qu\'imica-F\'isica, Universidad del Pa\' is Vasco, UPV/EHU, Leioa, Spain}
\affiliation{$^c$ IKERBASQUE, Basque Foundation for Science, E-48011 Bilbao, Spain}
\begin{abstract}
\noindent
{The place and role of an Observer in quantum mechanics has been a subject of an ongoing debate since the theory's inception. Wigner brought this question to the fore in a celebrated scenario in which a super-Observer observes a Friend making a measurement. Here we briefly review why this \e{Wigner Friend scenario} has been taken to require the introduction of the Observer's consciousness, or alternatively to show the inconsistency of quantum measurement theory. We will argue that quantum theory can consistently leave observers outside its narrative, by making only minimal assumptions about how the information about the observed results is stored in material records.}
\end{abstract}
\pacs{03.65.Ta, 03.65.AA, 03.65.UD}
\maketitle
\epigraph{ And I come to the fields and spacious palaces of my memory, where are the treasures of
innume- rable images, brought into it from things of all sorts perceived by the senses.
...What then we have utterly forgotten, though lost, we
cannot even seek after.} {St. Augustine of Hippo }
Are the rules of quantum mechanics internally consistent? To be consistent, the theory must give non-controversial 
prescriptions for any hypothetical situation, however difficult to realise in practice. An important question 
in this discussion is the extent to which quantum theory is able to deal with the role of a conscious Observer
when a measurement takes place. 
\newline
Some authors have considered consciousness to be instrumental in accounting for the observation of a measurement outcome. In their classic treatise, London and Bauer \cite{LB} described the state of an Observer's consciousness by a vector in a Hilbert space, that gets entangled with the system and the pointer in the pointer state basis. This cuts the von Neumann chain \cite{vN}, since consciousness gives an Observer the faculty to know his own state. Influenced by London and Bauer's work, Wigner \cite{Wig} described a scenario in which an agent needs to ascribe a state to another Observer undertaking a spin measurement. From the point of view of this agent, a \e{super-Observer}, unitary evolution would bring the Observer to absurd 
state of "suspended animation"\cite{Wig}. Consciousness is therefore assumed to be needed in modifying the linear laws of quantum evolution. Recently, several authors \cite{W1}-\cite{W3} used an extended Wigner's Friend scenario to question whether quantum mechanics rules have the necessary 
internal consistency.
\newline
In his Lectures, Feynman laid out the basis rules for evaluating probabilities with the help of probability amplitudes
defined for a sum over virtual paths \cite{FeynL}. 
The approach of  \cite{FeynL} is an agreement with Bohr \cite{Bohr} and von Neumann \cite{vN}  and leaves Observer's consciousness and sensations outside the theory's scope. This implies that the usual unitary evolution must be replaced by another type of evolution - state projection or \e{collapse} - when a measurement takes place. Otherwise the system simply gets entangled with the different components of the pointer, the environment etc., in a growing von Neumann chain \cite{vN}. 
\newline
The purpose of this paper is to review the original Wigner's Friend problem \cite{Wig} and, if possible, open up a new perspective on more complex situations, such as those considered in \cite{W1}-\cite{W3}.
\newline
The paper is organised as follows. In Sect. A we revisit standard quantum rules, as described in Feynman's text book \cite{FeynL}. Section B describes a setup in which a Friend(F) and Wigner (W) make their observations on a simple two-level systems, using their respective probes. 
In Sects. C to F we discuss possible choices open to F and W.
Section G briefly revisits Wigner's original analysis \cite{Wig}, and relates it to our own treatment of the problem.
Sections H and I contain the conclusions of our review, and discuss some of their general implications.
\section*{A. Classical vs. quantum rules} 
Classical mechanics predicts correlations between 
initial and final positions of the system, $q(t_0)$ and $q(t_1)$, as seen by an Observer \cite{Hertz}.
In its Hamilton's version,  one represent the system by a point, 
tracing a unique path in the phase space, the latter treated as a mathematical abstraction,
though reference to the physical world is secured by the equivalent Newtonian space-time formulation \cite{MATZ}. A system property is defined in terms of the phase-space variables and the property value depends on the given path irrespective of whether the property is observed or not.
The theory does not need, and hence makes no provision to account for the Observer's consciousness.
\newline
Quantum mechanics  also predicts correlations between two or more observations made on a quantum system. Its basic rules 
 were given in \cite{FeynL}, and we briefly review them here in a slightly tailored version (for more detail see 
\cite{DSepl}) 
If $L$ quantities $\mathcal Q^\l$, $\l=1,2,...,L$ are measured at different times $t=t_l$, one looks for a probabilities
$P(Q^L_{i_L}\gets Q^{L-1}_{i_{L-1}}....\gets Q^1_{i_1})$ to obtains a series of outcomes $Q^{\l}_{i_{\l}}$.
Each quantity is represented by a Hermitian operator {$\hat\Q^\l=\sum_{i^\l=1}^N |q^\l_{i_\l}\ra Q^\l_{i_\l}\la q^\l_{i_\l}|$,
where $N$ is dimension of the system's Hilbert space} . 
A {\it virtual}  (Feynman) path $\{q^L_{n_L}...\gets q^2_{n_2}\gets q^1_{n_1}\}$, connecting the eigenstates $|q^\l_{i_\l}\ra$, 
is endowed with a probability amplitude 
\begin{eqnarray} \label{01}
A(q^L_{n_L}...\gets q^2_{n_2}\gets q^1_{n_1}) =
\la q^L_{n_L}|\u(t_L,t_{L-1})|q^{L-1}_{n_{L-1}}\ra\times \n
...\la q^3_{n_3}|\u(t_3,t_{2})|q^{2}_{n_{2}}\ra
\la q^2_{n_2}|\u(t_2,t_{1})|q^{1}_{n_{1}}\ra,
\end{eqnarray}
where $\u(t',t)$ is the system's evolution operator. Initial measurement (preparation) must define the initial state 
$|q^1_{i_1}\ra$ unambiguously, $Q^1_{i_1} \leftrightarrow |q^\l_{i_\l}\ra$. An amplitude for the observed sequence
 ({\it real path}) 
 $Q^L_{i_L}\gets Q^{L-1}_{i_{L-1}}....\gets Q^1_{i_1}$ is obtained by by adding the amplitudes (\ref{01})
 according to the degeneracies 
of the eigenvalues $Q^\l_{i_\l}$, $\l=1,2,...L-1$ ( virtual paths ending in different final states do not interfere \cite{FeynL}). This yields a \e {sum over paths} type equation
\begin{eqnarray} \label{02}
A(q^L_{n_L}...\gets Q^{\l}_{i_{\l}}...\gets Q^1_{i_1}) =
{\sum_{n_2,n_3,...,n_{L-1}=1}^N} \prod_{\l=2} ^{L-1}
\q\n
{\Delta\left (Q^\l_{i_\l}-\la q^\l_{n_\l}|\Q^\l|q^\l_{n_\l}\ra\right )}
A(q^L_{n_L}\gets q^{L-1}_{n_{L-1}}...\gets q^1_{i_1}),\q
\end{eqnarray}
where $\Delta(x-y)= 1$ if $x=y$, and $0$ otherwise, account for the observed outcomes. The desired probability is found by taking the absolute square
of the amplitudes, 
\begin{eqnarray} \label{03}
P(Q^L_{i_L}\gets Q^{L-1}_{i_{L-1}}....\gets Q^1_{i_1})=
\n
\sum_{n_L=1}^N \Delta\left (Q^L_{i_L}-\la q^L_{n_L}|\Q^L|q^L_{n_L}\ra\right ) 
\n
|A(q^L_{n_L}\gets Q^{L-1}_{i_{L-1}}...\gets Q^1_{i_1})|^2,
\end{eqnarray}
Note that the eigenstates $|q^\l_{i_L}\ra$ are determined by the measurements the Observer is planning 
to make, and not by the actual state of the evolving system. 
The probability in Eq.(\ref{03}) refers to the entire series of the planned observations \cite{FeynL}, and 
does not refer explicitly to the collapse of the wave function.
\newline
Unlike the classical theory,  quantum mechanics makes only statistical 
predictions for a real path $Q^L_{i_L}\gets Q^{L-1}_{i_{L-1}}....\gets Q^1_{i_1}$.
(One exception is the choice
$\Q^L =\u(t_\l,t_1)\Q^\l\u^{-1}(t_\l,t_1)$, in which case an outcome, corresponding to the 
eigenvalue $\Q^\l_{i_\l}$
 may be obtained with certainty \cite{DSann}). 
Also, adding intermediate observations may destroy the interference between the Feynman paths in Eq.(\ref{02}), 
and different choices of measurements lead to essentially different statistical ensembles \cite{DSann}.
\newline
Classical mechanics usually makes no claims about 
living matter \cite{Hertz}.  In the original Wigner's Friend scenario \cite{Wig},
 Observer's consciousness is itself a subject of quantum mechanical description \cite{Wig1}.
Making it such a subject, however, is likely to lead to contradictions as we now discuss. 
\section*{B. A Wigner, his Friend and a spin}
Anticipating on Wigner's Friend scenario, to be introduced below, let us consider a system consisting of a spin-$1/2$, and two probes, one for the \enquote{Friend} (F), and one for the other observer (W).
The probes are two-level systems, which W and F can access directly, while the spin itself remains invisible to the naked eye. As described in the previous Section, their observations amount to  measuring projectors
\begin{eqnarray} \label{e1}
\hat Q^{F,W}=|1^{F,W}\ra \la 1^{F,W}|
\end{eqnarray}
with the  eigenvalues $Q_i^{F,W}$  $0$ and $1$. 
That is to say that F and W may 
\enquote{obtain outcomes} from
their respective probes, and see \enquote{yes} ($Q_1^{F,W}=1$) or \enquote{no} ($Q_1^{F,W}=0$) answers.
\newline
W and F are free to couple their probes to the spin, and to each other, if desired.
The outcome of an initial preparation ascertains that, at $t=t_0$, the probes and the spin can be described by a product state 
\begin{eqnarray} \label{e2}
|\Phi_0\ra= |0^W\ra |0^F\ra |s_0\ra.
\end{eqnarray}
We recall that  the probabilities in Eq.(\ref{03}) must refer to the sequences of the outcomes {\it experienced}, or {\it registered}  by the Observers \cite{vN}.
Thus, one may ask about the odds on F seeing a \enquote{yes}, and W seeing his  \enquote{yes}
later. Moreover, F can decide not to register his result. 
Then, by the strict rules imposed in the previous Section, one may only ask 
about the likelihood of W seeing a \enquote{yes}. The question is whether the rules of Sect. A allow a mere act of \enquote{registering a result} on the part of F can alter W's future experiences.
\section*{C. Both F and W couple their probes to the spin}
For simplicity, we assume that neither the probes, nor the spin 
 have their own dynamics, and remain in the same condition, unless F or W does something to them.
Let F briefly (practically instantaneously) couple his probe a $\tau^F$, 
 $ \tau^F>t_0$, and then look, or not look, at his result at $t_1> \tau^F$. The interaction (e.g., application of a CNOT gate) entangles the spin 
in {\color{black} an arbitrary} given state $|s\ra$ with F's probe according to 
 \begin{eqnarray} \label{f1}
|0^F\ra |s\ra \to \la s_1^F|s\ra  |1^F\ra|s_1^F\ra +\la s_2^F|s\ra |0^F\ra |s_2^F\ra, \q
\end{eqnarray}
 At $\tau^W > t_1$ Wigner couples his probe to the spin, using a different spin 
 basis $|s^W_i\ra$, 
  \begin{eqnarray} \label{f2}
|0^W\ra |s\ra \to \la s_1^W|s\ra |1^W\ra|s_1^W\ra +\la s_2^W|s\ra|0^W\ra |s_2^W\ra.
\end{eqnarray} 
The experiment is completed when W looks at his probe at $t=t_2 > \tau^W$. 
\newline 
The probability of W seeing a \enquote{yes} outcome (eigenvalue $1$, the probe state $|1^W\ra$) is  $P(yes^W)$. 
Would $P(yes^W)$ depend on whether F actually registered his result, or just  
turned on the coupling between his probe and the  the spin? In principle it could, since F's 
experience belongs to the past and, according to the rule of Sect.A, could destroy interference 
between virtual paths, leading to W's outcome \cite{DSepl}. In practice, it doesn't since there is no interference to destroy. 
The situation is sketched in Fig.1.  There are only four virtual paths with non-vanishing amplitudes, $\{m\}$, $m=1,2,3,4$,
  \begin{eqnarray} \label{f3}
A_1\equiv A(1' \gets 1 \gets \Phi_0) = \la1'|\hat U(\tau^W)| 1 \ra \la  1 |\hat U(\tau^F)|\Phi_0\ra\n
A_2\equiv A(2' \gets 1 \gets \Phi_0) ={\la2'|}\hat U(\tau^W)| 1 \ra \la  1 |\hat U(\tau^F)|\Phi_0\ra\n
A_3\equiv A(3' \gets 2 \gets \Phi_0) = \la3'|\hat U(\tau^W)| 2 \ra \la  2 |\hat U(\tau^F)|\Phi_0\ra\n
A_4\equiv A(4' \gets 2 \gets \Phi_0) = \la4'|\hat U(\tau^W)| 2 \ra \la  2 |\hat U(\tau^F)|\Phi_0\ra\n
\end{eqnarray} 
where  $\hat U(\tau^W)$ and $\hat U(\tau^F)$ are the evolution operators, corresponding to the 
evolutions (\ref{f1}) and (\ref{f2}), respectively, and we used a shorthand
  \begin{eqnarray} \label{f3a}
|1\ra \equiv | 0^W\ra |1^F\ra |s_1^F \ra,\q  |2\ra \equiv | 0^W\ra |0^F \ra |s_2^F \ra,\q \n
|1'\ra \equiv | 1^W\ra |1^F\ra |s^W_1 \ra, \q |2'\ra \equiv | 0^W\ra |1^F \ra |s^W_2 \ra,\n
|3'\ra \equiv | 1^W\ra |0^F\ra |s^W_1 \ra, \q |4'\ra \equiv | 0^W\ra |0^F \ra |s^W_2 \ra.
\end{eqnarray} 
The
 amplitudes in Eq.(\ref{f3}) can be expressed via the amplitudes, defined for the spin, uncoupled 
 {to} the probes,
  \begin{eqnarray} \label{f4}
A_1=\la s^W_1|s_1^F\ra\la s_1^F|s_0\ra, \q A_2=\la s^W_2|s_1^F\ra\la s_1^F|s_0\ra, \n
A_3=\la s^W_1|s_2^F\ra\la s_2^F|s_0\ra, \q A_4={\la s^W_2|s_2^F\ra}\la s_2^F|s_0\ra,
\end{eqnarray}  
The paths are shown in Fig.1a and, with
\begin{figure}[h]
\includegraphics[angle=0,width=8.8cm, height= 5cm]{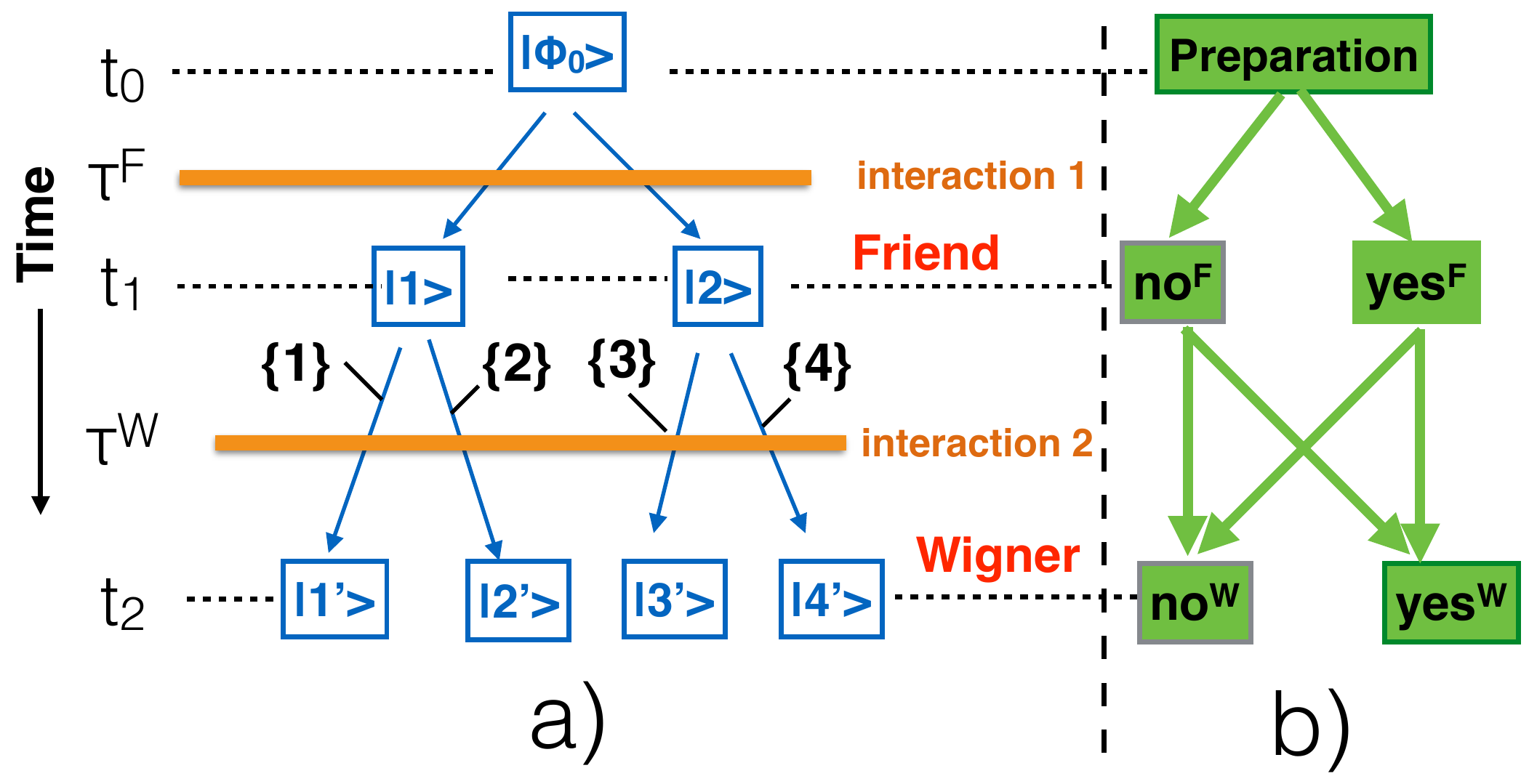}
\caption {a) Virtual paths in case F and W measure the {\it spin}. Interactions (\ref{f1}) and (\ref{f2})
at $\tau^F$ and  $\tau^W$ determine the evolution operators in Eq.(\ref{01}).
F and W register their outcomes at $t_1$ and $t_2$, respectively. The corresponding operators $\hat Q^\l$
in Eq.(\ref{02}) are the projectors (\ref{e1}). b) Sequences of observed events (real paths)
in case both F and W decide to register outcomes. W's probabilities do not depend on F's 
decision to register or not to register, as no interference between virtual paths is destroyed.}
\label{fig:FIG1}
\end{figure}
 both F and W looking at their results (see  Fig.1b),
we have 
  \begin{eqnarray} \label{f5}
P(yes^W,yes^F)=|A_1|^2,\q P(no^W,yes^F)=|A_2|^2,\n 
P(yes^W,no^F)=|A_3|^2,\q P(no^W,no^F)=|A_4|^2.\q
\end{eqnarray}
To evaluate W's probabilities in the case F was not registering his outcome, 
we must add amplitudes of all virtual paths, leading to the same final state $|i'\ra$.  However, there is  only one virtual path connecting $| \Phi_0\ra$ with each of the W's final states, so there is
 nothing to add. (Note that Fig.1a sketches an uninteresting double-slit problem, where four final positions on the screen, 
 $|1'\ra,..|4'\ra$,  can be reached through one of the slits, $|1\ra$ or $|2\ra$, only.) Therefore, from (\ref{03}) we have
   \begin{eqnarray} \label{f5}
P(yes^W|\text{F not registering})=|A_1|^2+|A_3|^2=\q\q\n
P(yes^W,yes^F)+
P(yes^W,no^F)\equiv P(yes^W|\text{F registering}),
\end{eqnarray}
 It does not matter whether F has registered his result or not, provided the probe was coupled to the spin. This 
 is an expected result, 
 to which we will return to it after considering 
first another example. 
\section*{D. W measures F's probe instead} 
Suppose that at $t=\tau^W$ Wigner decides to couple his probe not the spin, as before, but to his 
Friend's probe instead.  For this purpose, he uses a different basis ($i=1,2$)
  \begin{eqnarray} \label{g1}
|\phi_i^F\ra =u_{i1}|1^F\ra+u_{i0}|0^F\ra, \q \la \phi^F_i|\phi^F_j\ra =\delta_{ij},
\end{eqnarray} 
and with F's probe initially in some $|\psi^F\ra$ the coupling produces an entangled state  \begin{eqnarray} \label{g2}
|0^W\ra |\psi^F\ra \to \la \phi^F_1|\psi^F\ra |1^W\ra|\phi^F_1\ra +
\la \phi^F_2|\psi^F\ra |0^W\ra|\phi^F_2\ra.
\end{eqnarray} 
The situation is still described by the diagram in Fig.1a,  
with the only difference that now
  \begin{eqnarray} \label{f3b}
|1'\ra \equiv | 1^W\ra |\phi_1^F\ra |s_1^F \ra, \q |2'\ra \equiv | 0^W\ra |\phi_2^F \ra |s_1^F \ra,\q\n
|3'\ra \equiv | 1^W\ra |\phi_1^F\ra |s_2^F \ra, \q |4'\ra \equiv | 0^W\ra |\phi_2^F \ra |s_2^F \ra.\q
\end{eqnarray} 
The four possible outcomes are still as shown in Fig.1b, the probabilities are given by Eq.(\ref{f5})  and, as before, F's decision to register or not to register his outcome does not change the statistics of the results experienced by W. Next we discuss the reason for that.
\section*{E. The \enquote{in principle} principle. Von Neumann chains}
In his lectures \cite{FeynL} Feynman stressed that scenarios which can be distinguished 
{\it in principle} cannot interfere. Thus, it should not matter whether a conscious Observer has actually experienced a
particular outcome, as long as any Observer could experience it, perhaps at a later time.
\newline
In the example of Sect. C, F's observation finds the spin in one of its states $|s^F_i\ra$,
The spin's condition changes when W applies his coupling.
However, W's manipulation does not affect F's probe, which continues to carry the record of spin's 
condition, as it was just after $t_1$. Wigner's Friend
may decide to observe his probe later, or not to register his result at all (see Sect. 3-2 of \cite{FeynL}), and this is enough 
to preclude interference. {With F's machine switched off, the paths $\{1^W 0^Fs^W_1 \gets 0^W 0^F s_1^F\gets \Phi_0\}$
and $\{1^W 0^Fs^W_1 \gets 0^W 0^F s_2^F\gets \Phi_0\}$ would interfere, but with the machine on, 
the paths $\{1^W1^F s^W_1 \gets 0^W 1^F s_1^F\gets \Phi_0\}$
and $\{1^W0^F s^W_1 \gets 0^W0'^Fs_2^F\gets \Phi_0\}$ lead to different final states, and become {\it exclusive 
alternatives} \cite{FeynL}.}
\newline
Similarly in Sect.D W destroys the state of F's probe, but leaves alone the spin itself.
Therefore, F can repeat his measurement using a different probe at  a $t_1 > t_2$, and obtain the same 
result he would have seen if he had bothered to look at some $t_1 < t_2$. This information is now encoded in the 
spin's, rather than in the probe's condition.
\newline
In practice, Friend's states $|1^F\ra$  and $|0^F\ra$
(and similarly $|1^W\ra$  and $|0^W\ra$) may describe not a single degree of freedom, but a sequence 
of $K$ objects and devices, starting with a simple pointer, coupled to the spin,  passing through an amplifier to  Friend's retina and neurons, and ending 
at the elusive boundary, where information about physical world enters the Observer's \enquote{extra observational inner life} \cite{vN}.
With all elements of the chain in agreement, we can write
  \begin{eqnarray} \label{f3b}
|0^F\ra =\prod_{k=1}^K |0_k^F\ra, \q|0^F\ra|s_1^F\ra =\prod_{k=1}^K |0_k^F\ra|s_1^F\ra.
\end{eqnarray}
 and then group the terms in an arbitrary manner. For example it is possible to redefine
 $|0^F\ra \to \prod_{k=n}^K |0_k^F\ra$ and $|s_1^F\ra \to \prod_{k=1}^{n}|0_k^F\ra|s_1^F\ra$, 
 or even $|0^F\ra \to \prod_{k=m}^n |0_k^F\ra$ and $|s_1^F\ra \to \prod_{k=1}^{m}|0_k^F\ra
 \prod_{k=n}^{K}|0_k^F\ra |s_1^F\ra$.
Wigner is free to couple his probe to the newly defined $|0^F\ra$ or $|s_1^F\ra$ as before, and with the
same result. For as long as a single $|0_k^F\ra$ or $|s_1^F\ra$ remains to carry the evidence of what 
F {would} have seen at $t=t_1$, actually seeing it has no effect on W's experience. 
We haven't however yet discussed the original Wigner's argument, and will do it next. 
\section*{F. The Wigner's Friend problem}
In a nutshell, the problem discussed in \cite{Wig} concerns the case where W decides to engage the entire composite {\it F's probe + spin}, so that no material record of what happened at $t_1$ is carried forward 
for future reference. In order to do so he may couple his probe  at $t=\tau^W$ thus 
entangling it with a composite's state $|\varphi\ra$ 
 \begin{eqnarray} \label{h1}
|0^W\ra |\varphi \ra \to \la 1^{FS}|\varphi \ra |1^W\ra|1^{FS}\ra +\n
\sum_{i=2}^4\la i^{FS}|\varphi \ra |0^W\ra |i^{FS}\ra, \q
\end{eqnarray}
using, for example,
 \begin{eqnarray} \label{h2}
|1^{FS}\ra= \left [|1^F\ra |s_1^F\ra+ |0^F\ra |s_2^F\ra\right ]/\sqrt{2},\n |2^{FS}\ra = [|1^F\ra |s_1^F\ra- |0^F\ra |s_2^F\ra]/\sqrt{2}.
\end{eqnarray}
{(Two remaining orthogonal basis states, $|3^{FS}\ra$ and $|4^{FS}\ra$, are not connected by the evolution operators in (\ref{01}),
and need not be specified.)} 
Two possible final states of the composite
{\it W's probe + F's probe + spin}, therefore, are
 \begin{eqnarray} \label{h2}
|1'\ra= |1^W\ra|1^{FS}\ra , \q \text{and} \q |2'\ra= |0^W\ra|2^{FS}\ra.es
\end{eqnarray}
Now the four virtual paths, shown in Fig.2a, correspond 
to a primitive double-slit problem with only two final positions, $|1'\ra$ and $|2'\ra$, each of which is accessible via both \enquote{slits} $|1\ra$ and $|2\ra$.
\begin{figure}[h]
\includegraphics[angle=0,width=8.8cm, height= 4.8cm]{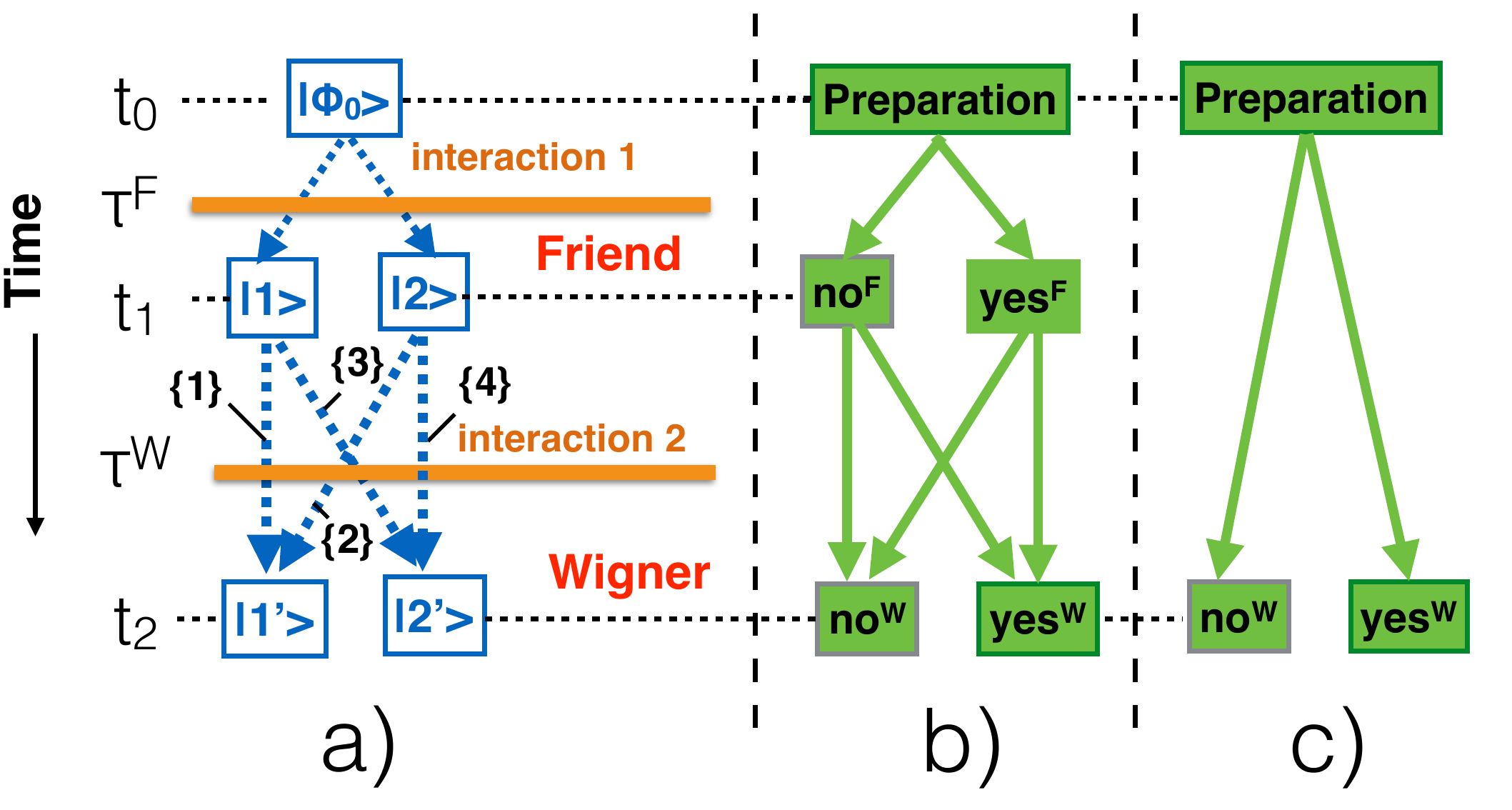}
\caption {a) Virtual paths in case F measures the {\it spin}, and W measures
 a composite {\it spin + F's probe}. The evolution operators in Eq.(\ref{01}) are determined 
 by interactions (\ref{f1}) and (\ref{h1}), and the operators $\hat Q^\l$ are still the projectors (\ref{e1}).
 b) Real paths in case both F and W look at their probes. c) Real paths if F decides not to look.
According to the rules of Sect. A, W's probabilities depend on F's decision. The \enquote{paradox} disappears 
if F's looking leaves a record in his memory, which adds an extra degree(s) of freedom to W's calculation.}
\label{fig:FIG1}
\end{figure}
The corresponding amplitudes are given by
{\color{black}
\begin{eqnarray} \label{h3}
A_1=\la1'|U(\tau^W)|1\ra\la1|U(\tau^F)|\Phi_0\ra,\n A_2=\la1'|U(\tau^W)|2\ra\la2|U(\tau^F)|\Phi_0\ra,\n
A_3=\la2'|U(\tau^W)|1\ra\la1|U(\tau^F)|\Phi_0\ra,\n A_4=\la2'|U(\tau^W)|2\ra\la2|U(\tau^F)|\Phi_0\ra.
\end{eqnarray}}
Both interactions (\ref{f1}) and (\ref{h1}) are now in place. The question is whether F registering his outcome 
at $t=t_1$ would change the odds on W obtaining  his \enquote{yes} outcome at $t=t_2$. Our rules of Sect.A
 say that it would. Indeed, with F {\it registering}, there are four real observable outcomes shown in Fig.2b, 
and 
\begin{eqnarray} \label{h4}
P(yes^W|\text{F registering})=|A_1|^2+|A_2|^2=\n
{P(yes^W,yes^F)}+P(yes^W,no^F)\q\q\q
\end{eqnarray}
while with F {\it not registering}, we  have a different result
\begin{eqnarray} \label{h5}
P(yes^W|\text{F not registering})=|A_1+A_2|^2 =\n
P(yes^W|\text{F registering}) + \text{Re}[A_1^*A_2].\q\q\q
\end{eqnarray}
This is the result we want to examine, but first we briefly revisit Wigner's own argument.
\section*{G. Wigner's treatment of the problem}
Wigner`s formulation of the dilemma was slightly different. He asked (in our notations) whether
just after F registered his result  at $t=t_1+0$, W should ascribe to the composite {\it F's probe + spin}
a pure state {
\begin{eqnarray} \label{j1}
|\phi\ra =\la s_1^F|s_0\ra |1^F\ra|s_1^F\ra + \la s_2^F|s_0\ra |0^F\ra|s_2^F\ra,
\end{eqnarray}
or a statistical mixture
\begin{eqnarray} \label{j2}
\hat\rho  = |1^F\ra|s_1^F\ra|\la s_1^F|s_0\ra|^2\la s_1^F| \la1^F| +\n
 |0^F\ra|s_2^F\ra|\la s_2^F|s_0\ra|^2\la s_2^F| \la0^F|.\q
\end{eqnarray}}
This is not the language we used in Sect.A.  Feynman's rule does not refer directly to the state 
of the system at a given time, or to the \enquote{collapse of the wave function} following an observation. 
\newline	
Translated into the language of the previous Section, Wigner's choice is  between considering  the paths $\{1\}$ and $\{2\}$
(as well as $\{3\}$ and $\{4\}$) in Fig.2a as interfering or as exclusive alternatives. Indeed, if  (\ref{j1}) is true, 
the probability of W seeing a \enquote{yes} is given by 
\begin{eqnarray} \label{j3}
P(yes^W)\equiv \text{tr}\left [|1'\ra \la 1'|\otimes \hat\rho \right ]=|A_1|^2 + |A_2|^2.
\end{eqnarray}
If, on the other hand, (\ref{j2}) holds, the same probability should be 
\begin{eqnarray} \label{j4}
P(yes^W)\equiv |\la 1'|\phi\ra|^2=|A_1 + A_2|^2.
\end{eqnarray}
Feynman's prescription, however, is clear. With both F and W looking, interference must be destroyed, 
and Eq.(\ref{j2}) rather than Eq.(\ref{j1}) must be used. 
As far as we know, while discussing distinguishable scenarios \cite{FeynL}, 
Feynman never specified how they can be distinguished and by who.
\newline
Wigner, for his part,  brought this question to the fore, and made three points . 
Firstly, a superposition (\ref{j1}) is allowed for inanimate objects (including macroscopic ones), but ought to be considered absurd if applied to a conscious F, thereby forced to remain in a state of \enquote{suspended animation} until W asks him what he saw. 
\newline
Secondly, and because of that, F's consciousness should act onto material objects, so that the wave function of 
{\it F+F's probe+spin} is turned into a statistical mixture. 
\newline
Thirdly,  in practice, telling the difference between a pure state and 
a mixture can be extremely difficult for sufficiently complex systems, such as F and his macroscopic probe (laboratory). 
\newline
 Now much depends on how F's consciousness and quantum theory are interrelated.
 According to London and Bauer \cite{LB}, F's consciousness should be described by a quantum state, which we could include in the Friend's states denoted $|0^F\ra$ or $|1^F\ra$ in Eq. (\ref{h1}).
This would have to destroy the superposition (\ref{h1}) because, by introspection, consciousness knows that a single outcome has been observed. But standard linear quantum mechanics lacks the means for doing so. 
In \cite{Wig} Wigner provides the necessary means postulating a non-linear evolution whenever human 
consciousness is involved. Indeed, if standard linear QM were to be applied , the composite {\it F consciousness + F's probe} would be in a superposition that Wigner asserts \e{is not credible} \cite{Wig2}.
Next we turn to Ref. \cite{FeynL} for more insight.
\section*{H. Feynman's rules and material records }
In our attempt to follow the approach of \cite{FeynL} and Sect. A,  we would need to observe at least the following four restrictions.
\newline
(i) The probabilities in Eq.(\ref{03}) refer to the impressions, registered by conscious Observers,
 (see also \cite{vN}).
\newline
(ii) Quantum theory, has nothing to say about consciousness itself, or about its interaction 
with material (i.e., inanimate) world. 
\newline
(iii) Quantum theory, [i.e., Eqs.(\ref{01})-(\ref{03})],  applies to all material  objects, regardless of their size and
complexity.
\newline
(iv) (Uncertainty Principle) In the Young's double-slit experiment it is impossible 
 to know which slit was
chosen by a particle, while  maintaining the interference pattern on the screen  \cite{FeynL}. 
\newline
The second assumption appears to be the most vulnerable, and we are forced at least to speculate 
about some properties of Observer's consciousness and the existence of material records. 
\newline
We begin with a contradiction.  If a record of a registered outcome is kept inside F's \enquote{inner world}
(which we can say nothing about), and W manages to entangle with his probe all objects holding the material records [but not F's consciousness (see ii)], 
we contradict the Uncertainty Principle (iv). Indeed, W detects the presence of the interference term in Eq.(\ref{h4}), 
and yet F {\it knows} that the system has passed through the \enquote{slit}, represented by, say, $|1\ra$ in Fig.2a. 
\newline
An attempt to remedy this by including F's inner world into a quantum mechanical calculation, as suggested by London and Bauer \cite{LB}  returns 
us to the unsavoury notion of the \enquote{state of suspended animation} \cite{Wig}, experienced by F prior to W's query.
\newline
A possible way out requires making certain minimal assumptions about the existence of material records.
The rules of Sect. A readily account for the discrepancies between probabilities in Eqs.(\ref{h4}) and (\ref{h5}), 
if each act in which an Observer registers his/her result is accompanied by producing a record 
in the memory, or indeed on any other material object. Adding the record's degree(s) of freedom would 
change the size of a Hilbert space and create new virtual and real paths. Thus, two calculations of Sect. F, one for F registering his result, and the other for F not doing so, would naturally 
be different. 
\newline
Now by (iii) we must assume that the record left in F's memory or elsewhere is, in principle, accessible
to W's manipulations. There is, however, no contradiction.
If all material records could be erased by W's subsequent measurement, previous F's experience would not count
-it could be undone and would never be confirmed- and W's probabilities would be given by Eq.(\ref{h5}). If W's measurement misses at least one material record left
in F's memory or in his laboratory, the presence of the corresponding orthogonal states will guarantee that 
Eq.(\ref{03}) will yield the probabilities (\ref{h4}). Subsequent observations on the part of F cannot change this 
result, since the rules (\ref{02}) and (\ref{03}) are explicitly causal, and forbid the influence of future measurements 
on the results already obtained  (see, for example, \cite{DSepl}). 
{\color{black} In effect, we are able to leave aside the very act of registering, and concentrate on its physical consequences.}
Quantum mechanics only briefly looses the narrative while the perceived outcome is being passed 
to the Observer's memory, but quickly recovers it after a tangible material evidence is provided. 

Finally, it is worth recalling, that quantum theory can be used (and is most often used) by a third person, say, W's cousin (C), who is reasoning about the 
joint experiences of F and W, and is not  taking part in the actual experiments herself. Her conclusion must be 
that the odds on W seeing a \enquote{yes} outcome are given by Eq.(\ref{h4}) if his measurement preserves some form of F's record, 
and by Eq.(\ref{h5}) if W completely destroys it. Personalities do not matter, and in C's mind  particular F and W
can be replaced by any pair of two human Observers, which may or may not communicate with each other
during the course of the experiment.
\section*{I. Discussion and conclusions }
Quantum theory is "invented" (in the words of Wigner \cite{Wig}) by human consciousness for dealing with material physical phenomena and, from the start, we do not expect it to be a suitable judge of  the consciousness itself \cite{ADB}.
For this to be true one needs  to show that the theory remains internally consistent, if restricted to physical phenomena only.
This idea arises already in classical physics \cite{Hertz}, and is exacerbated in the quantum case, 
where the theory predicts essentially different statistical ensembles, with and without the presence of an intermediate 
Observers.  Wigner's radical answer to this conundrum \cite{Wig} was to bring the Observer's consciousness into the theory's remit in another way, by modifying the standard (linear) quantum mechanical evolution.
One may reasonably wonder whether the problem can be treated without such a dramatic departure from conventional quantum theory
\newline
With this in mind, we have argued that Feynman's rules, as given in \cite{FeynL}, allow one
to leave conscious Observers outside the theory's scope, on a condition that all information about the outcomes 
of the observations made be contained in material records, themselves subject to a quantum analysis. 
Certain, albeit minimal, assumptions about the Observer's behaviour are, therefore, necessary.
In particular, whenever asked about the outcome of an experiment, 
the Observer would need to consult the relevant record, and cannot simply be "aware of it" at all times. Such a record 
can be kept as a note on a piece of paper, a file on a microchip, or in the Observer's own memory. 
\newline
The rules of Sect.A 
automatically account for the existence of such a record, by including the corresponding  degree of freedom
into a quantum mechanical calculation. Virtual paths, previously leading to the same final state, 
are thus modified to lead to distinguishable outcomes.
This is sufficient for destroying the interference  present if  a Hilbert space of a smaller dimension is used, as happens, for example, in  Fig.1. 
\newline
A record can be destroyed, e.g., by subsequent measurements, 
as happens in Fig.2. Although some records, e.g., macroscopic ones may be more robust, 
all information can be destroyed in this manner in principle, if not in practice. 
With all records destroyed,  one must conclude that the information about the 
outcome of a particular experiment is irretrievably lost, in a stronger sense than in the classical case.
{\color{black} It was argued by Feynman and Hibbs \cite{FeynH}, 
by design a measurement apparatus yields a stable record in situations in which, through the statistical mechanics of amplification, the amplitudes play no more role. In \cite{FeynL} Feynman argues also that even if a photon, scattered 
each time an electron passes through the first slit of the Young's experiment, is never observed, the interference pattern will be destroyed. In both cases we deal with records robust in the sense that in practice they 
will never be engaged by the participants of the experiment, which one would take into account while 
evaluating the probabilities of the outcomes.}
\newline
Finally, allowing the Observer to keep information about his/her perceived outcomes in a domain beyond the reach of quantum theory would lead to a conflict with the Uncertainty Principle. Indeed, in Fig.2 the Principle demands that the outcome of F's measurement 
remain {\it indeterminate}, since the paths $\{1\}$ and $\{3\}$ ($\{2\}$ and $\{4\}$) interfere. 
Friend's awareness of his result would, in this case, amount to knowing the slit, chosen by an electron in the Young's 
double-slit experiment {\it and} maintaing the interference pattern on the screen. Given the Principle's crucial role in "protecting" quantum mechanics from  a logical collapse, \cite{FeynL}, avoiding such conflict should be a necessary requirement for any analysis of the Wigner's Friend problem. 
 \begin{center}
\textbf{Acknowledgements}
\end{center}
Financial support of
MCIU, through the grant
PGC2018-101355-B-100(MCIU/AEI/FEDER,UE)  and the Basque Government Grant No IT986-16.
is acknowledged by DS.

  \end{document}